# Evaluation of dynamic fracture toughness of a bonded bi-material interface subject to high-strain-rate shearing using digital image correlation

**Tomohisa Kojima, Yuta Kimura, Shuichi Arikawa, Mitsuo Notomi**

**Highlights**

· Dynamic shear test with three-layered bonded bi-material test piece.

· Location of the interface crack tip was identified with digital image correlation.

· Determining the interfacial fracture toughness with digital image correlation.

**Evaluation of dynamic fracture toughness of a bonded bi-material interface subject to high-strain-rate shearing using digital image correlation**

**Author names and affiliations**

**Tomohisa Kojima[1]  (corresponding author)**

Department of Mechanical Engineering, School of Science and Technology, Meiji University

1-1-1, Higashimita, Tama-ku, Kawasaki-shi, Kanagawa, 214-8551, Japan

kojima.31k@g.chuo-u.ac.jp

**Yuta Kimura**

Department of Mechanical Engineering Informatics, Graduate School of Science and Technology, Meiji University

1-1-1, Higashimita, Tama-ku, Kawasaki-shi, Kanagawa, 214-8551, Japan

**Shuichi Arikawa**

Department of Mechanical Engineering Informatics, School of Science and Technology, Meiji University

1-1-1, Higashimita, Tama-ku, Kawasaki-shi, Kanagawa, 214-8551, Japan

**Mitsuo Notomi**

Department of Mechanical Engineering, School of Science and Technology, Meiji University

1-1-1, Higashimita, Tama-ku, Kawasaki-shi, Kanagawa, 214-8551, Japan

**Abstract**

   The fracture toughness at the bi-material interface of layered test pieces is determined using digital image correlation (DIC) measurements as part of dynamic shear tests. High-strain-rate shear tests are conducted on a three-layered bonded test piece comprising a central aluminum layer with PMMA resin layers. When the crack is sufficiently shorter than the bonding interface length, mode II fracture appears at the aluminum–PMMA interface; otherwise, mode I and mode II fractures appear. As the obtained stress intensity factor is consistent with the literature values, a reasonable interface fracture toughness can be evaluated using the DIC.

---

[1] Present address: Department of Precision Mechanics, Faculty of Science and Engineering, Chuo University

1-13-27, Kasuga, Bunkyo-ku, Tokyo, 112-8551, Japan

**Nomenclature**

| | |
|---|---|
| $C$ | correlation coefficient |
| $E$ | Young's modulus |
| $F$ | luminance value at the image coordinates before deformation |
| $G$ | luminance value at the image coordinates after deformation |
| $K$ | stress intensity factor |
| $K_0$ | overall stress intensity factor |
| $K_\mathrm{I}$ | mode I stress intensity factor |
| $K_\mathrm{II}$ | mode II stress intensity factor |
| $r$ | distance from the crack tip |
| $t$ | time |
| $u$ | displacement |
| $x$ and $y$ | image coordinates before deformation |
| $x^*$ and $y^*$ | image coordinates after deformation |
| $\varepsilon$ | strain |
| $\varepsilon'$ | measured strain by digital image correlation |
| $\gamma, \eta, \kappa$ | bi-material constant |
| $\mu$ | shear modulus |
| $\nu$ | Poisson's ratio |
| $\sigma_{yy}$ | stress in the $y$-direction |
| $\sigma_{xy}$ | shear stress |
| $\sigma_{\theta\theta}$ | tangential stress |
| $\sigma_{r\theta}$ | shear stress |
| if | interface |
| al | aluminum |
| ac | PMMA |
| $x$ | $x$-direction |
| $y$ | $y$-direction |
| $xy$ | shear |
| DIC | digital image correlation |
| PMMA | poly (methyl methacrylate) |
| SHPB | split Hopkinson pressure bar |

## 1 Introduction

The strengths of adhesive bonded and composite materials with dissimilar interfaces are greatly influenced by the interfacial strength. As products and equipment composed of these materials are often subjected to impact (such as in the cases of car accidents or bird strikes), the dynamic interfacial strength of a bi-material is a critical parameter that must be evaluated, for which several methods currently exist. According to one study investigating interfacial strength, over 200 such evaluation methods exist [1]. However, most of these methods essentially involve quasi-static tests to assess structural performance, of which only two are applicable to dynamic interfacial strength evaluation. These are the Charpy impact test and the impact wedge-peel test [2]. Though these test methods have been specified as part of ASTM standards [3], Japanese Industrial Standards [4] and standards of the International Organization for Standardization [5], they are expedient industrial test methods wherein the basic intrinsic strength of the adhesive cannot be determined because of complex stress distributions therein.

When a material with dissimilar interfaces is subjected to impact, the propagation of stress waves imposes a complicated loading, which is different from quasi-static loading. Here, we consider a material with a dissimilar interface, to which an impact load is applied, having elastic wave propagation velocities $c_1$ and $c_2$ in a direction parallel to the interface, as shown in Fig. 1. Because the speeds of propagation of the elastic stress waves differ, the displacement of materials also differs; consequently, shear deformation occurs at the interface. Ultimately, the shear deformation increases over time and the stress waves are reflected repeatedly in the material, resulting in a complicated force being applied to the interface.

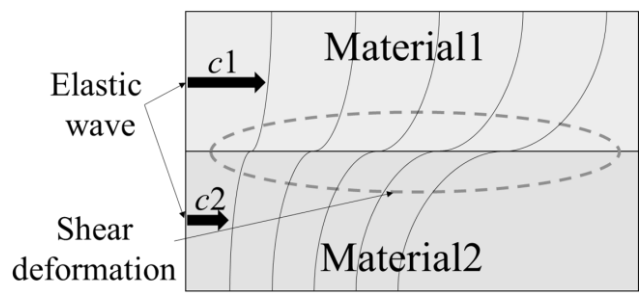

Figure 1. Illustration of shear deformation on a bi-material interface with stress wave propagation

Numerous studies, including many analytical studies, have investigated fracture parameters of bi-material interfaces for many years [6]–[8]. The fracture parameters [9,10], the effects of elastic and plastic properties [11,12], and displacement and strain fields within the interfacial region [13] have also been analyzed. A recent concern arising from these investigations is regarding mixed-mode fracture parameters for dissimilar interfaces in thin layers [14]–[17]. However, not many studies have investigated dynamic interfacial fractures [18,19]. In parallel with these studies, many experimental studies have also been conducted using optical techniques to visualize the fracture behavior of the interface and to determine various fracture parameters [20]–[22]. In particular, photoelasticity has been widely employed in such studies [23]–[26]. Other studies have employed quasi-static loading [27]–[29], investigated crack extension under dynamic loading conditions [30]–[32], and identified fracture criteria [33]–[35].

In contrast, the digital image correlation (DIC) method [36]–[38] has only been investigated to a greater extent in recent years because of the growing interest in optical techniques. DIC is an optical measurement technique that involves the examination of displacement upon a comparison of a random pattern drawn on the surface of the object before and after deformation. This technique does not require a complicated optical system and can enable measurement of the displacement and strain of the entire field, as well as the usage of the photoelastic method. DIC can also be employed in the case of non-transparent materials wherein the photoelastic method cannot be applied. Regarding materials with dissimilar interfaces, various studies have determined crack tip opening displacements [39,40], identified cracks [41], and determined mixed-mode stress intensity factors based on the displacement distributions near the crack tips [42] and indirect measurements of ultra-high-speed shear strain at the interfaces [43]. Studies investigating the interface of composites have combined the split Hopkinson pressure bar (SHPB) and mesh-DIC methods, where dynamic displacements and strain distributions in the composites were photographically measured using high-speed video cameras [44]. However, despite the investigations of all the aforementioned studies, we find that a method for directly determining the fracture toughness of the interface from the measurement result of DIC has not yet been proposed.

Accordingly, this study aims to apply the SHPB method to determine the fracture toughness values at the interface via observation of crack propagation in a high-strain rate shear test. The study employs test pieces of bonded materials composed by combining materials with different elastic wave propagation velocity characteristics. The propagation behavior of elastic stress waves in the test pieces was observed using DIC and an ultrahigh-speed video camera, and the fracture toughness was evaluated.

## 2 Methods

### 2.1 Experimental apparatus

Figure 2 shows a schematic of the experimental apparatus employed in this study, including the combination of the SHPB method and high-speed imaging equipment. For using the SHPB method, phosphor bronze 16 mm in diameter was employed for the striker, incident bar, and output bar, which were 300, 1800, and 500 mm in length, respectively. Two strain gauges were attached at each strain measurement point. The output was subsequently amplified. The bending strain was eliminated via the formation of a bridge circuit. The strain rate during the test was $-5.1 \times 10^3$ s$^{-1}$. To perform high-speed imaging, a high-speed camera and a laser light source were synchronized using the output from the strain gauge—which is attached to the incident bar—as a trigger signal. An ultra-high-speed video camera (Kirana, Specialised Imaging Ltd.) at a frame rate of 1,000,000 fps and a 200-W laser light source (CAVILUX Smart, Cavitar Ltd.) of 640 nm wavelength were used.

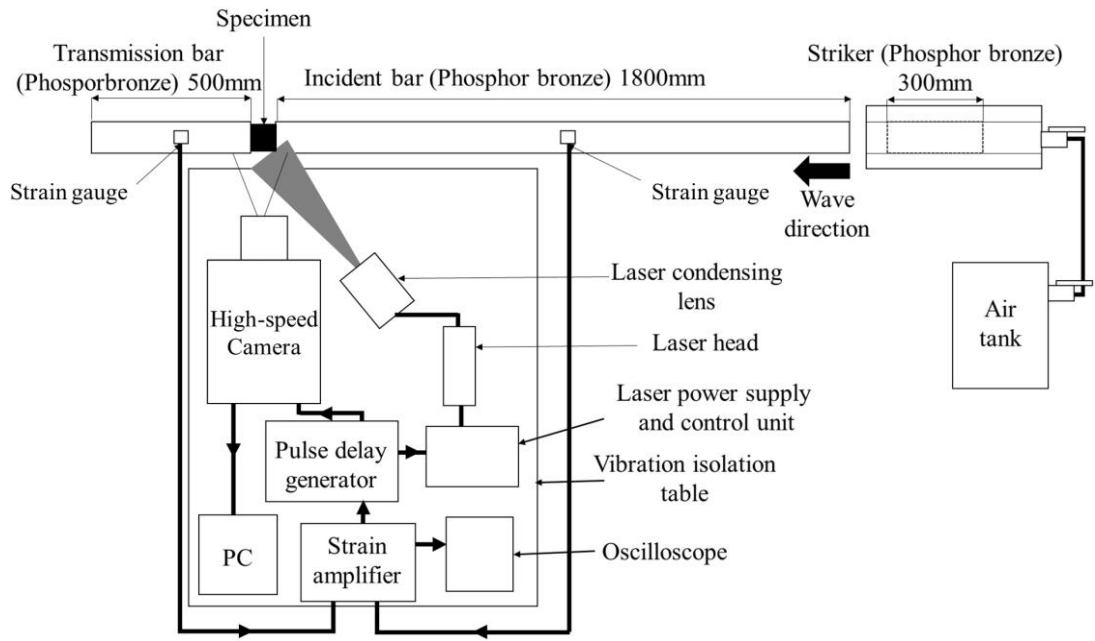

Figure 2. Schematic of the experimental apparatus

## 2.2 Measurement principles of DIC

Figure 3 shows a conceptual representation of DIC. The displacement is determined by using Eq. 1 in a process of searching for a region, called a subset from the transformed image. This process exploits the fact that there is no change in the random pattern of the object surface before and after the deformation of the object. Although the texture of the surface may change as the structure deforms, the subset deforms accordingly. In this study, bilinear interpolation was used as a luminance-value interpolation method for detecting displacements on a subpixel basis, and the Newton–Raphson method was used as an iterative computation algorithm.

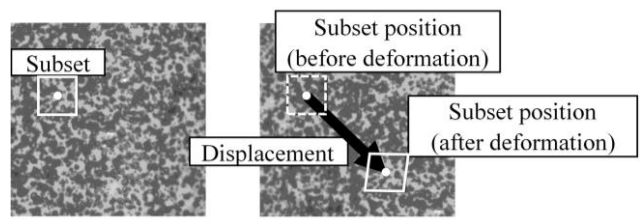

Figure 3. Principle of DIC

$$C(x, y, x^*, y^*) = \frac{\sum F(x,y) G(x^*, y^*)}{\sqrt{\sum F(x,y)^2 \sum G(x^*, y^*)^2}} \quad (1)$$

where *C* is the correlation coefficient, *F* is the luminance value at the image coordinates (*x*, *y*) before deformation, and *G* is the luminance value at the image coordinates (*x**, *y**) after deformation.

**2.3 Test piece and DIC conditions**

The test piece employed in this study is presented in Fig. 4. It was fabricated by bonding an aluminum alloy (A6063) bar of size 25 mm × 5 mm × 5 mm with two poly(methyl methacrylate) (PMMA) resin bars of the same dimensions on either side (Al/PMMA test piece). The material properties of aluminum [45] and PMMA [46] are listed in Table 1. In general, the speeds of sound (longitudinal wave speed) of aluminum and PMMA are approximately 6,400 m/s and 2,700 m/s, respectively. As can be seen, the speeds of sound in each material are significantly different. To confirm the influence of the adhesive layer, another test piece was prepared using PMMA for all three layers (PMMA test piece). By the fracture pattern of the PMMA test piece, it was attempted to verify if the bonding adhesive layer could be regarded as part of the PMMA bar in the test piece or not. The bonding positions were shifted by 5 mm, as shown in Fig. 4, and only the aluminum section was brought into contact with the incident bar and PMMA with the output bar. Thus, the length of the bonding interface was 20 mm, and the total length was 30 mm.

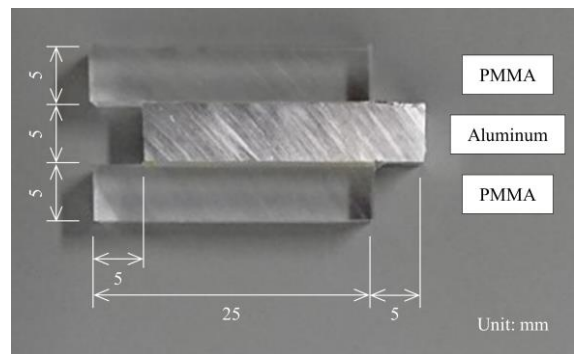

Figure 4. Bonded Al/PMMA layered test piece (PMMA /A6063/PMMA)

Random patterns were applied to the surfaces of these test pieces for the DIC analysis. The surface of the test piece was polished with #400 emery paper, and then the random pattern was painted so that it would not be peeled off even in the impact test. First, a thin coating of white was applied as the base with a spray. Next, after it dried, the pattern was created with a black spray. An acrylic spray and a silicon acrylic spray were used for white and black, respectively. For the DIC analysis, the pattern spot size (average diameter) should correspond to a few pixels at the imaging plane of the camera. It was reported that the absolute error of the DIC analysis is minimum in the speckle size range of 4–5 pixels [47]. The size of a single pixel of the high-speed camera used in the study was 30 μm × 30 μm. Besides, the optical magnification of the lens was set to 1; hence, we set the target spot size to 150 μm, which is equivalent to 5 pixels. Afterwards, we chose a black spray that easily creates spots equivalent to 3–8 pixels.

The DIC analysis was conducted after the tests to obtain the displacements and strain distribution. The

4Accepted manuscriptDIC conditions of the subset size, strain gauge length, and interpolation method were set to 31 × 31 pixels, 31 pixels, and bi-linear, respectively. The recorded video frames were of a resolution of 60 μm/pixel. The PMMA test piece and the Al/PMMA test piece were then subjected to three experiments each.

Table 1. Material properties [45,46]

| Material | Density (kg/m$^3$) | Young's modulus (GPa) | Poisson's ratio |
|---|---|---|---|
| Aluminum (A6063) | 2,700 | 70 | 0.33 |
| PMMA | 1,200 | 3.0 | 0.35 |

### 2.4 Stress intensity factor at the bi-material interface

If the interfacial strength is significantly lower than the strength of the bonded material, crack development at the interface can be evaluated based on the stress intensity factor. The dynamic stress displacement field of the crack tip in the homogeneous material obtained considering inertia is similar to that obtained in the static case, except that the stress intensity factor $K_I(t)$ depends on time $t$ [48,49]. The situation is also considered to be the same in the case of a crack along with an interface. A method for determining the stress intensity factor of cracks on the interface of dissimilar materials is shown below [8]. As the normal stress and the shear stress involve vigorous oscillation at different phases near the crack tip on the dissimilar material interface, the crack development is evaluated based on the overall stress intensity factor. Assuming a planar stress field and considering only two-dimensional deformation, the overall stress intensity factor $K_0$ can be expressed as per Eq. (2).

$$K_0 = \sqrt{K_I^2 + K_{II}^2} = \frac{\sqrt{2\pi r}}{\cosh(\pi \eta)} \left[ \sigma_{\theta\theta}^2 + \sigma_{r\theta}^2 \right]^{1/2} \quad (2)$$

Here, $\eta$ is the bi-material coefficient and can be expressed as per the following equation.

$$\eta = \frac{1}{2\pi} \ln \gamma \quad (3)$$

and

$$\gamma = \frac{\mu_1 + \kappa_1 \mu_2}{\mu_2 + \kappa_2 \mu_1} \quad (4)$$

where $\mu$ is the shear modulus, $\kappa = (3 - 4v)/(1 + v)$ in the planar stress state, and $v$ is Poisson's ratio. Subscripts 1 and 2 correspond to two different materials. Because the stress intensity factor on the interface may be set to $\theta = 0$ in the equation above, the following equation (Eq. (5)) is obtained. Here, the direction parallel to the interface was defined as the $x$-direction, and the direction perpendicular to the interface was defined as the $y$-direction.

$$K_0 = \frac{\sqrt{2\pi r}}{\cosh(\pi \eta)} \left[ \sigma_{yy}^2 + \sigma_{xy}^2 \right]^{1/2} \quad (5)$$

### 2.5 Stress-strain conversion at the bi-material interface using DIC results

To calculate the stress intensity factor at the interface, the stress at the interface was calculated using the Please cite this article as: T. Kojima, Y. Kimura, S. Arikawa, M. Notomi, Evaluation of dynamic fracture toughness of a bonded bi-material interface subject to high-strain-rate shearing using digital image correlation, *Engineering Fracture Mechanics*, Vol. 241 (2021), doi: https://doi.org/10.1016/j.engfracmech.2020.107391.

obtained strain distributions in the *x*- and *y*-directions as well as the shear strain. For materials containing interfaces such as the bi-material interface in this case, when a stress wave propagates in a direction parallel to the interface, the strain in the direction parallel to the interface becomes continuous on the interface. On the contrary, though the stress is continuous, the strain is discontinuous in the direction perpendicular to the interface, as illustrated in Fig. 5. However, the strain distribution calculated via DIC is smoothened by the influence of the subset range at the time of calculation of the displacement. The smoothening is also due to the influence of the gauge length at the time of the strain calculation. As a result, the strain distribution in the vicinity of the interface becomes continuous. Calculation of stress using a continuous strain results in stress discontinuity being shown to occur at the interface. Therefore, the obtained strain values need to be revised. Regarding the measurement results of DIC, the length affected by the subset and gauge length is (subset size + gauge length)/2. In this study, because both the subset size and gauge length are set as 31 pixels, the affected range also corresponds to 31 pixels.

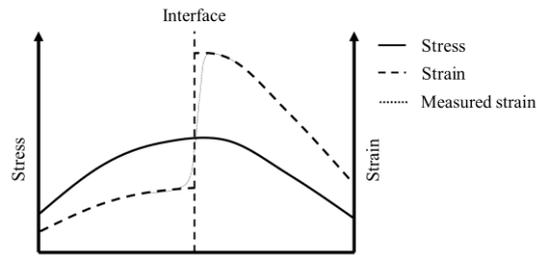

Figure 5. Distribution of stress and strain near the interface

The displacement of aluminum and PMMA influences the displacement field in the vicinity of the interface by the same ratio, as obtained using DIC. Because a similar effect applies to the calculation of strain, the value of strain at the center of the interface can be obtained by averaging the values of strain on both sides. Therefore, the strain of the interface as measured by DIC is expressed as follows,

$$\begin{cases} \varepsilon'_{\mathrm{ifx}} = \dfrac{\varepsilon_{\mathrm{alx}} + \varepsilon_{\mathrm{acx}}}{2} \\ \varepsilon'_{\mathrm{ify}} = \dfrac{\varepsilon_{\mathrm{aly}} + \varepsilon_{\mathrm{acy}}}{2} \end{cases} \qquad (6)$$

Here, the subscripts 'al', 'ac', and 'if' represent aluminum, PMMA, and the interface, respectively. Assuming a planar stress field, the *x*- and *y*-directional strains of aluminum and PMMA can be expressed as

$$\begin{cases} \varepsilon_{alx} = \dfrac{1}{E_{al}}\left(\sigma_{alx} - \nu_{al}\sigma_{aly}\right) \\ \varepsilon_{aly} = \dfrac{1}{E_{al}}\left(\sigma_{aly} - \nu_{al}\sigma_{alx}\right) \\ \varepsilon_{acx} = \dfrac{1}{E_{ac}}\left(\sigma_{acx} - \nu_{ac}\sigma_{acy}\right) \\ \varepsilon_{acy} = \dfrac{1}{E_{ac}}\left(\sigma_{acy} - \nu_{ac}\sigma_{acx}\right) \end{cases} \quad (7)$$

Because of the continuity on the interface, we have

$$\begin{cases} \varepsilon_{ifx} = \varepsilon_{alx} = \varepsilon_{acx} \\ \sigma_{ify} = \sigma_{aly} = \sigma_{acy} \end{cases} \quad (8)$$

Substituting Eqs. (7) and (8) into Eq. (6) makes it possible to obtain $\sigma_{ify}$ as follows:

$$\sigma_{ify} = \dfrac{E_{al}E_{ac}}{(1-\nu_{ac})E_{al} + (1-\nu_{al})E_{ac}}\left[(\nu_{al}+\nu_{ac})\varepsilon'_{ifx} + 2\varepsilon'_{ify}\right] \quad (9)$$

Next, the shear strain at the interface as measured by DIC can be expressed as

$$\varepsilon'_{ifxy} = \dfrac{\varepsilon_{alxy} + \varepsilon_{acxy}}{2} \quad (10)$$

and the stress at the interface is given by

$$\sigma_{ifxy} = \mu_{al}\varepsilon_{alxy} = \mu_{ac}\varepsilon_{acxy} \quad (11)$$

From Eqs. (10) and (11), we obtain

$$\sigma_{ifxy} = \dfrac{2\varepsilon'_{ifxy}\mu_{al}\mu_{ac}}{\mu_{al} + \mu_{ac}} \quad (12)$$

From the above, the stress intensity factor at the interface can be obtained from the result of the DICs using Eqs. (5), (9), and (12).

## 3 Experimental results

### 3.1 Fracture pattern of the PMMA test pieces

Figure 6 shows the condition of the PMMA test pieces after the test was completed. All the test pieces contained a portion where the interface was broken and a portion where the center PMMA bar was broken, as can be seen in the figure. Based on such a fracture pattern, it may be confirmed that, in the three-layer test piece employed in this study, the strength of the bonding interface was not lower than that of the bonded PMMA resin. Therefore, the test piece could be regarded as an integral PMMA material including the bonding layer.

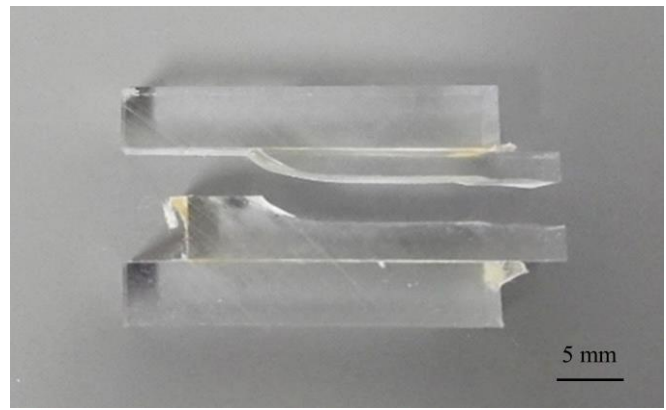

Figure 6. PMMA test piece after the test

### 3.2 Fracture pattern of the PMMA/Al test pieces

The photographs of the PMMA/Al test piece during the experiment are shown in Fig. 7. The time shown below the images is the time since the stress wave reached the test piece. Figure 7 shows the occurrence of cracks from the left on the upper interface, which is indicated as the white line. It was observed that the white line penetrated the entire interface at 52 $\mu$ s, then the interface peeled off and the test piece began to fall while rotating at 64 $\mu$ s.

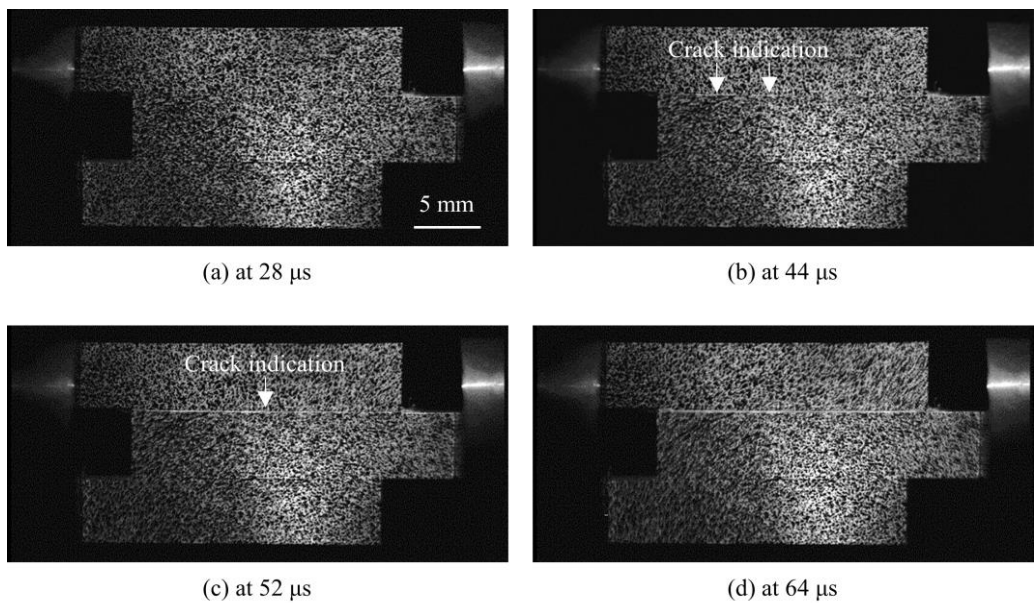

(a) at 28 µs            (b) at 44 µs

(c) at 52 µs            (d) at 64 µs

Figure 7. Fracture process of the Al/PMMA test piece (a) at 28 µs, (b) at 44 µs, (c) at 52 µs, and (d) at 64 µs

Figure 8 shows the Al/PMMA test piece after the test. The fracture occurred at the interface between aluminum and one of the PMMA bars, and the adhesive was completely retained on the PMMA side after the fracture.

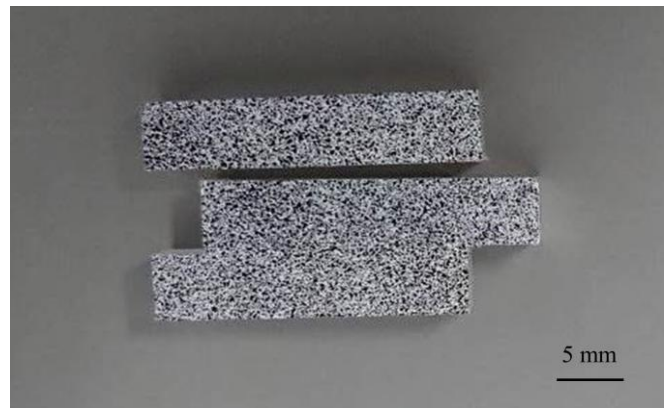

Figure 8. Fractured Al/PMMA test piece after the test

**3.3 Displacement and strain distributions of the Al/PMMA test piece**

Figures 9–13 show the results of the DICs. The displacement distributions in the *x*- and *y*-directions, strain distributions of in the *x*- and *y*-directions, and shear strain distribution are shown in numerical order in the figures. In Figs. 9 and 10, it can be seen that the stress waves first propagated to the aluminum section of the test piece and then gradually propagated to the PMMA section; thus, increasing displacement with time was exhibited. The displacements became discontinuous at the upper interface 38 μs after the test was commenced, and those at 52 μs after the test was commenced became discontinuous over the entire interface. In this manner, the occurrence of cracks and delamination at the interface could be confirmed based on the displacement distribution measured using DIC.

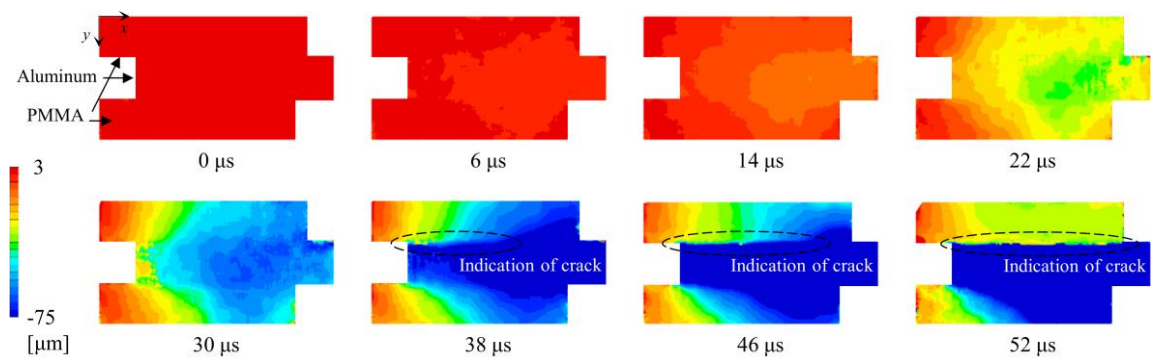

Figure 9. Distribution of the displacement in the *x*-direction ($u_x$)

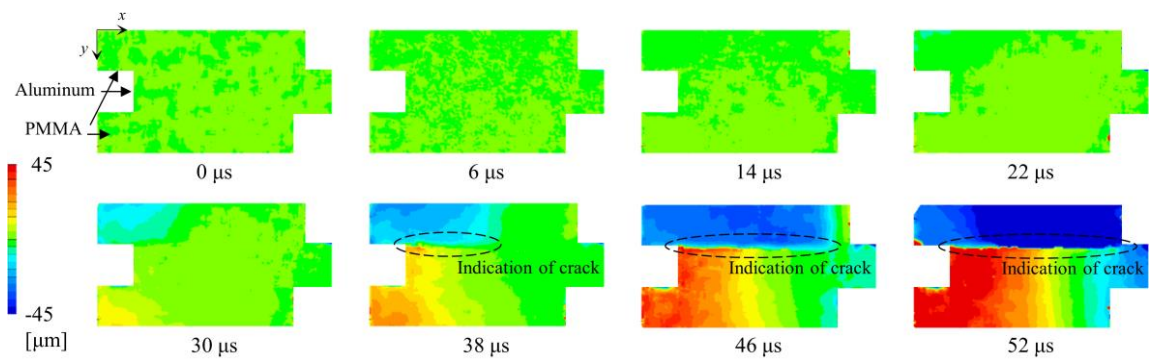

Figure 10. Distribution of the displacement in the $y$-direction ($u_y$)

From Figs. 11 and 12, it can be seen that although the disturbance was large, the strain was generated only towards the left end of the test piece. It appears that the strain generated in the PMMA section was larger than that in the aluminum section. Figure 11 suggests that the crack is indicated by the appearance of the $x$-direction compressive strain and the strain value of 0 alternately on the upper interface 38 μs after the commencement of the test. As shown in Fig. 12, the crack opening is indicated by the occurrence of $y$-direction strain observed at the rear end of the upper interface. From these displacement and strain distributions, an observation could not be made regarding the distribution owing to the existence of a yielding area in the vicinity of the crack tip. Therefore, it appeared that the yielded zone at the crack tip was sufficiently small to apply the small-scale yielding approximation. Moreover, it is indicated that the interfacial strength of the Al/PMMA interface was considerably low in the experiment. The shear strain did not increase from the rear end of the upper interface; instead, it increased simultaneously on the entire interface (Fig. 13).

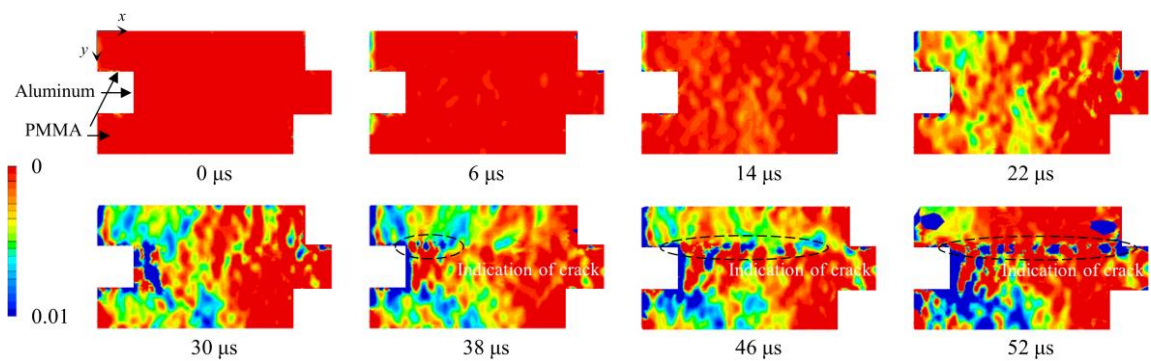

Figure 11. Distribution of the strain in the $x$-direction ($\varepsilon_x$)

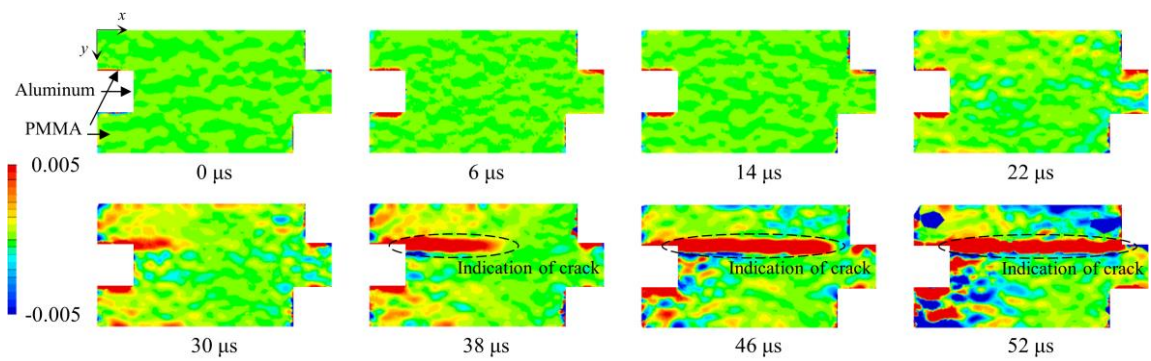

Figure 12. Distribution of the strain in the *y*-direction ($\varepsilon_y$)

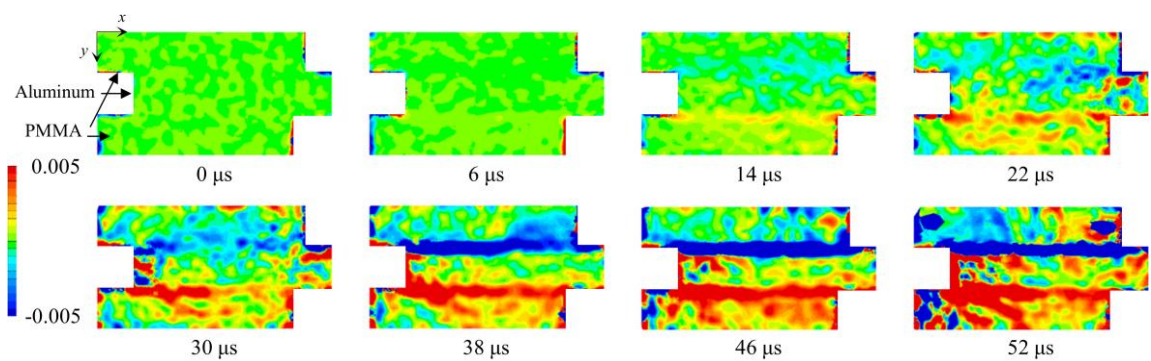

Figure 13. Distribution of the shear strain ($\varepsilon_{xy}$)

**4 Evaluation of the fracture toughness at the bi-material interface**

**4.1 Identification of the location of the crack tip**

To estimate the fracture toughness value of the bi-material interface, we attempted to determine the location of the crack tip. Here, the DIC results were used to determine the location. As discussed in Section 3.2, the presence of cracks causes a discontinuity of the displacement at the interface (Figs. 9–10). Therefore, the crack tip would exist at a location where the displacement on the interface becomes discontinuous. We adopted the *x*-direction displacement to locate the crack tip because the *y*-directional displacement also involves the influence of the Poisson effect brought about as a result of the compression of the test piece. Figure 14 shows an example of the location of a discontinuity in the *x*-direction displacement on the interface.

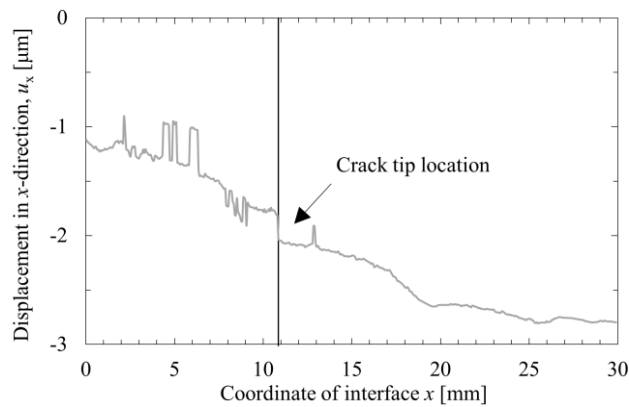

Figure 14. Location of a discontinuity in the x-direction displacement on the interface

**4.2 Stress and stress intensity factor at the interface**

The equation derived in the previous section was employed to calculate the stress and the stress intensity factor at the interface. The calculated interfacial stresses, $\sigma_{ify}$ and $\sigma_{ifxy}$, are shown in Fig. 15. The location of the edge of the test piece that comes into contact with the output bar was set to zero in the horizontal axis shown in Fig. 15. The stresses were calculated at 35, 40, 42, and 45 μs after the stress wave reached the test piece. The crack lengths at each point of time were 11.0, 12.9, 15.9, and 18.3 mm, respectively. From Fig. 15, it can be seen that the shear stress has a constant value at the interface. On the contrary, the stress in the $y$-direction is not constant but corresponds to a wavy curve, as shown. This is because disturbance exists owing to an error in the displacement field obtained via DIC. In the strain field obtained upon differentiating the displacement field, a large disturbance occurs due to an increase in the error.

The value of stress in the $y$-direction is large for the section of the test piece where the interface is peeled off on the left side of the crack tip. This is because the opening of the crack causes displacement, in turn resulting in an apparent strain. In Figs. 15 (a) and (b), a wavy curve is seen, implying unsteady values, but the value of the $y$-directional stress on the right side of the crack tip appears to be more constant. However, it was not constant, as can be seen in Figs. 15 (c) and (d). This suggests that a moment was applied in the vicinity of the crack tip owing to the influence of the progression of the crack and the reduction in the bonding area; resultantly, a fracture of mode I as well as mode II, appeared. Thus, mode II fracture may appear when the crack is short, as compared to the interface length, and mode I fracture may appear when the crack is longer in comparison.

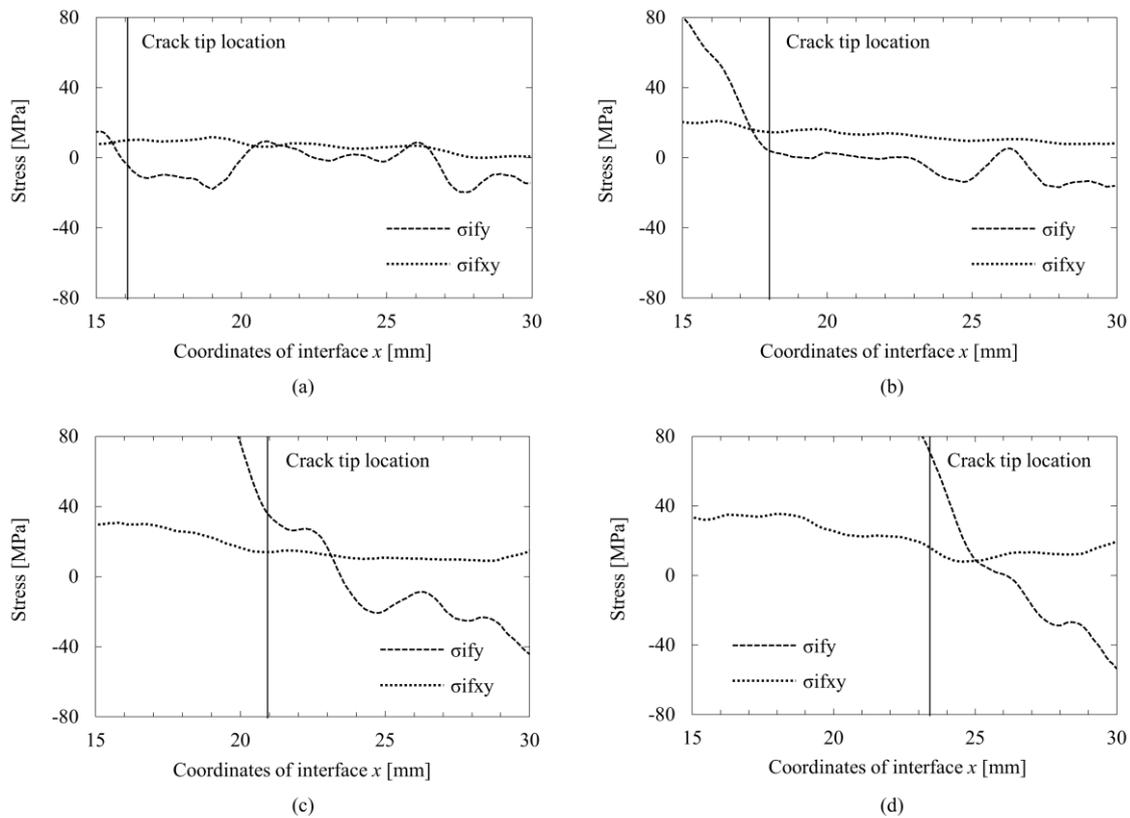

Figure 15. Stress distribution along the interface with the crack length and time of (a) 11.0 mm, at 35 μs, (b) 12.9 mm, at 40 μs, (c) 15.9 mm, at 42 μs, and (d) 18.3 mm, at 45 μs

The calculated stress intensity factors corresponding to the points along the interface are shown in Fig. 16. Here, it is understood that the value of $K_{0c}$ is disturbed by the influence of the disturbance of the $y$-direction stress. $K_{0c}$ becomes zero at the crack tip, where only a singular point exists. If the vibration component due to the error from the DIC result is removed, the value of $K_{0c}$ does not appear to be changed even for a location slightly distant from the crack tip.

Because the stress value includes the disturbance resulting from the DIC error, the average value of the stress intensity factor for the vicinity of the crack tip was calculated and compared (Fig. 17). The averages were calculated in four ranges, 70, 50, 40, and 30 pixels. The ratios to the subsets are 2.3, 1.6, 1.3, and 1, respectively, and correspond to lengths of 4.2, 3.0, 2.4, and 1.8 mm, respectively. As described earlier, when the crack length is greater than the joint length, a moment is applied. Thus, the stress condition at the interface is considered to be changed. As a result, the values of the stress intensity factor vary depending on the range of the calculated mean values of the crack length, i.e., 15.9 and 18.3 mm, as shown in Fig. 16. However, by changing the range in which the average value is calculated, it can be seen that the value of $K_{0c}$ assumes a similar value regardless of the crack length value.

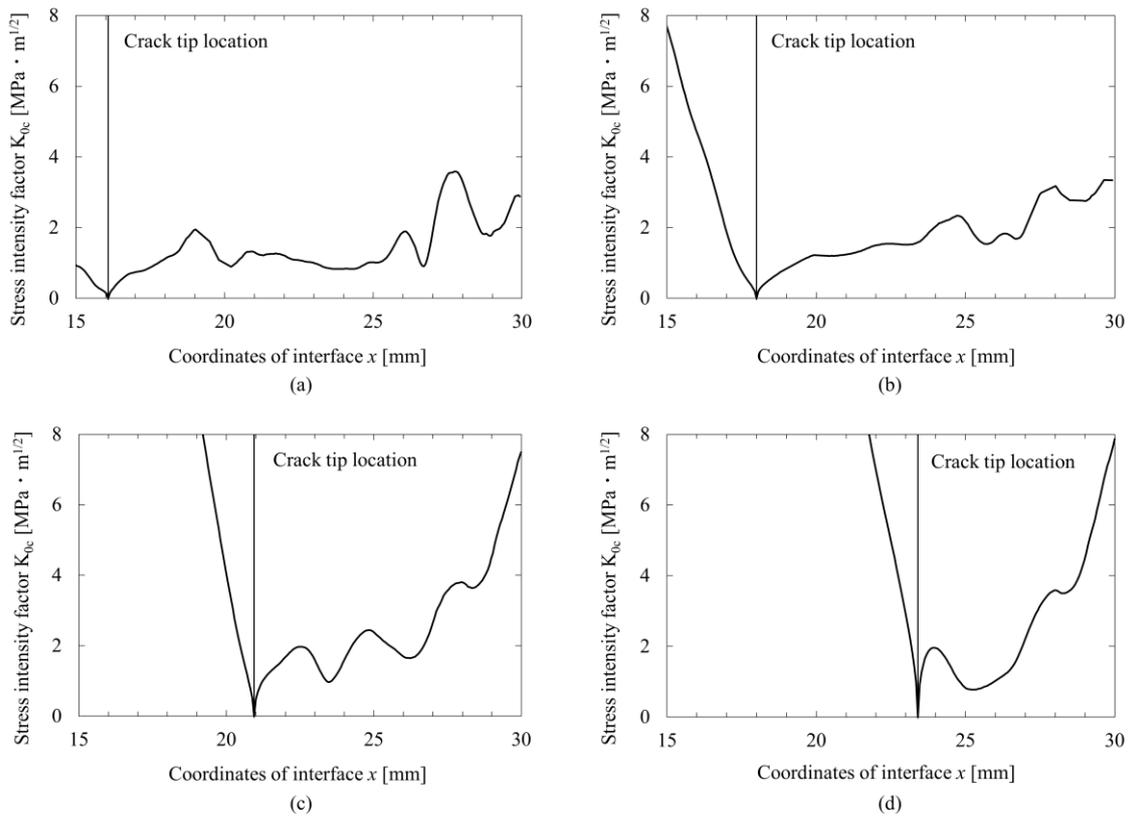

Figure 16. Stress intensity factor along the interface with crack lengths and times of (a) (a) 11.0 mm, at 35 μs, (b) 12.9 mm, at 40 μs, (c) 15.9 mm, at 42 μs, and (d) 18.3 mm, at 45 μs

Upon averaging the stress intensity factor values from the crack tips to the length ranges with a ratio of 1 to the subset, the obtained average value was 1.38 MPa. Tippur et al. experimentally determined the dynamic stress intensity factor with crack tip mixity at a bonded Al/PMMA interface to be 0.6–1.2 MPa・$m^{1/2}$ [32]. Nishimura et al. [50] reported that the values of the stress intensity factor for different interfaces in the case of mode II fracture were 1.2 MPa・$m^{1/2}$. These values were obtained upon conducting the bending test from the resin section of the test piece, to which the IC sealing resin and the lead frame material (Fe-42Ni) were bonded. The values are close to those obtained in the research cited earlier, although there are differences in terms of materials and test methods. Therefore, it may be observed that a reasonable stress intensity factor could be evaluated using the method proposed in this study.

We believe that the proposed method can be applied to other bi-materials. However, in the case of a metal-to-metal bi-material, as the deformation is small with metals, it may be difficult to identify the crack tip location. Therefore, when the proposed method is applied to metal–metal, it would be necessary to introduce a pre-crack and then magnify the area around the crack tip for obtaining images for the DIC analysis.

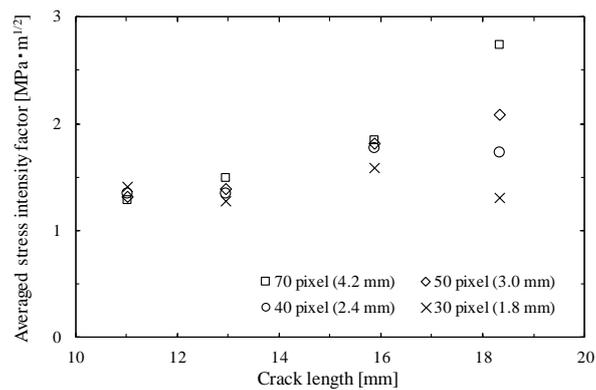

Figure 17. Summary of the average values of the stress intensity factor ($K_{0c}$) calculation range.

## 5 Conclusion

In this study, high-strain-rate shear tests were conducted on a three-layered bonded test piece comprising a central aluminum layer with PMMA resin layers bonded on both sides. Upon calculating the displacement field and the strain field using DIC, the crack tip was located, and the fracture toughness was evaluated at the bonded bi-material interfaces.

As a result of the DIC, using the PMMA/Al/PMMA test piece, it was possible to determine the process by which 1) the elastic stress wave propagated to the aluminum section, 2) the wave was transmitted to the PMMA section, and 3) the crack developed at the interface. The tip of the crack was identified using displacement distributions in the direction parallel to the interface by exploiting the fact that the displacement was not continuous at the location where the crack existed.

The fracture toughness of the interface was evaluated using the stress intensity factor. First, to calculate the stress intensity factor at the bonded bi-material interface, the true interface stress was calculated by correcting the strain value at the interface obtained using DIC. The distribution of the stress in the direction perpendicular to the interface suggested that mode II fracture appears in the present test method when the crack is sufficiently shorter than the length of the bonding interface, and mode I and mode II fractures appear in the test method when the crack is longer in comparison. Although the value of the stress intensity factor was disturbed by the error of the DIC analysis, it was confirmed that the obtained values of the stress intensity factor were similar regardless of the difference in the crack length, upon averaging the stress intensity factor values from the crack tips to the length range with a ratio of 1 to the subset. As the obtained stress intensity factor value was similar to the values calculated in the related literature, it can be concluded that the method proposed in this study yields a reasonable stress intensity factor.